\documentclass[twocolumn]{aastex62}

\usepackage{color}
\usepackage{xspace}
\usepackage{url}
\usepackage{graphicx}
\usepackage{subfigure}

\shorttitle{\textit{Fermi} GBM Observations of GRB 150101B}
\shortauthors{Burns et al.}

\begin{document}

\title{\textit{Fermi} GBM Observations of GRB 150101B:\\ A Second Nearby Event with a Short Hard Spike and a Soft Tail}

\correspondingauthor{Eric Burns}
\email{eric.burns@nasa.gov}

\author{E.~Burns}
\affiliation{NASA Postdoctoral Program Fellow, Goddard Space Flight Center, Greenbelt, MD 20771, USA}

\author{P.~Veres}
\affiliation{Center for Space Plasma and Aeronomic Research, University of Alabama in Huntsville, Huntsville, AL 35899, USA}

\author{V.~Connaughton}
\affiliation{Science and Technology Institute, Universities Space Research Association, Huntsville, AL 35805, USA}

\author{J.~Racusin}
\affiliation{NASA Goddard Space Flight Center, Greenbelt, MD 20771, USA}

\author{M.~S.~Briggs}
\affiliation{Space Science Department, University of Alabama in Huntsville, Huntsville, AL 35899, USA}
\affiliation{Center for Space Plasma and Aeronomic Research, University of Alabama in Huntsville, Huntsville, AL 35899, USA}

\author{N.~Christensen}
\affiliation{Physics and Astronomy, Carleton College, MN 55057, USA}
\affiliation{Artemis, Universit\'{e} C\^{o}te d'Azur, Observatoire C\^{o}te d'Azur, CNRS, CS 34229, F-06304 Nice Cedex 4, France}

\author{A.~Goldstein}
\affiliation{Science and Technology Institute, Universities Space Research Association, Huntsville, AL 35805, USA}

\author{R.~Hamburg}
\affiliation{Space Science Department, University of Alabama in Huntsville, Huntsville, AL 35899, USA}
\affiliation{Center for Space Plasma and Aeronomic Research, University of Alabama in Huntsville, Huntsville, AL 35899, USA}

\author{D.~Kocevski}
\affiliation{Astrophysics Branch, ST12, NASA/Marshall Space Flight Center, Huntsville, AL 35812, USA}

\author{J.~McEnery}
\affiliation{NASA Goddard Space Flight Center, Greenbelt, MD 20771, USA}

\author{E.~Bissaldi}
\affiliation{Istituto Nazionale di Fisica Nucleare, Sezione di Bari, I-70126 Bari, Italy}
\affiliation{Dipartimento Interateneo di Fisica, Politecnico di Bari, Via E. Orabona 4, 70125, Bari, Italy}

\author{T. Dal Canton}
\affiliation{NASA Postdoctoral Program Fellow, Goddard Space Flight Center, Greenbelt, MD 20771, USA}

\author{W.~H.~Cleveland}
\affiliation{Science and Technology Institute, Universities Space Research Association, Huntsville, AL 35805, USA}

\author{M.~H.~Gibby}
\affiliation{Jacobs Technology, Inc., Huntsville, AL 35805, USA}

\author{C.~M.~Hui}
\affiliation{Astrophysics Branch, ST12, NASA/Marshall Space Flight Center, Huntsville, AL 35812, USA}

\author{A.~von~Kienlin}
\affiliation{Max-Planck-Institut f\"{u}r extraterrestrische Physik, Giessenbachstrasse 1, 85748 Garching, Germany}

\author{B.~Mailyan}
\affiliation{Center for Space Plasma and Aeronomic Research, University of Alabama in Huntsville, Huntsville, AL 35899, USA}

\author{W.~S.~Paciesas}
\affiliation{Science and Technology Institute, Universities Space Research Association, Huntsville, AL 35805, USA}

\author{O.~J.~Roberts}
\affiliation{Science and Technology Institute, Universities Space Research Association, Huntsville, AL 35805, USA}

\author{K. Siellez}
\affiliation{Center for Relativistic Astrophysics and School of Physics, Georgia Institute of Technology, Atlanta, GA 30332, USA}

\author{M.~Stanbro}
\affiliation{Space Science Department, University of Alabama in Huntsville, Huntsville, AL 35899, USA}

\author{C.~A.~Wilson-Hodge}
\affiliation{Astrophysics Branch, ST12, NASA/Marshall Space Flight Center, Huntsville, AL 35812, USA}

\begin{abstract}
In light of the joint multimessenger detection of a binary neutron star merger as the gamma-ray burst GRB 170817A and in gravitational waves as GW170817, we reanalyze the \textit{Fermi} Gamma-ray Burst Monitor data of one of the closest short gamma-ray bursts: GRB 150101B. We find this burst is composed of a short hard spike followed by a comparatively long soft tail. This apparent two-component nature is phenomenologically similar to that of GRB 170817A. While GRB 170817A was distinct from the previously known population of short gamma-ray bursts in terms of its prompt intrinsic energetics, GRB 150101B is not. Despite these differences, GRB 150101B can be modeled as a more on-axis version of GRB 170817A. Identifying a similar signature in two of the closest short gamma-ray bursts suggests the soft tail is common, but generally undetectable in more distant events. If so, it will be possible to identify nearby short gamma-ray bursts from the prompt gamma-ray emission alone, aiding the search for kilonovae.
%gravitational waves from neutron star mergers and kilonova.
\end{abstract}

% %% Keywords should appear after the \end{abstract} command. 
% %% See the online documentation for the full list of available subject
% %% keywords and the rules for their use.
% \keywords{(stars:) gamma-ray burst: individual (GRB 150101B)}%, (stars:) gamma-ray burst: general, notices --- 
% % miscellaneous --- catalogs --- surveys}

\section{Introduction} \label{sec:intro}

\citet{Burns_GBM_BAT} asserted that extremely close short gamma-ray bursts (SGRBs) are not necessarily bright. The assertion is based on the lack of correlation between redshift and the prompt SGRB brightness at Earth and the modest fluence (compared to other SGRBs) of the nearby short GRB 150101B. This was 
%against conventional wisdom, but 
potentially very important given the new era of gravitational wave (GW) astronomy and the unique science possible only with joint GW-GRB detections. This expectation was spectacularly confirmed by the joint multimessenger detection of the merging of two neutron stars in gravitational waves as GW170817 \citep{GW170817-GW} by Advanced LIGO \citep{AdvLIGO} and Advanced Virgo \citep{AdvVirgo} and in gamma-rays as the low-luminosity GRB 170817A \citep{GBM_only_paper,INTEGRAL_only_paper} by the \textit{Fermi} Gamma-ray Burst Monitor (GBM; \citealt{FermiGBM}) and by the SPectrometer on-board INTEGRAL Anti-Coincidence Shield (SPI-ACS; \citealt{SPIACS}). The results from the joint detection confirmed binary neutron star mergers as progenitors of SGRBs and measured the speed of gravity to within one part in 10$^{-15}$ of the speed of light \citep{GW170817-GRB170817A}. To maximize the science from multimessenger observations of neutron star mergers we need to identify nearby events and fully understand them. GRB 170817A falls within the normal SGRB distributions of fluence, peak flux, peak energy, and duration for SGRBs 
\citep{GBM_only_paper}. However, it appears to be composed of a short hard spike (similar to the usual phenomenology of more distant SGRBs) followed by a longer soft tail that may be thermal in origin, falls on the softer and longer ends of the hardness and duration distributions of SGRBs (as observed with GBM) \citep{GBM_only_paper}, and is several orders of magnitude less luminous than other SGRBs with known redshift \citep{GW170817-GRB170817A}.

We now return to GRB 150101B. New analysis on fine timescales of this burst, presented here, uncovers a short hard spike followed by a longer soft tail that may be thermal in origin. As GRBs 170817A and 150101B are among the closest SGRBs (of those with unambiguously measured redshifts, see \citealt{GW170817-GRB170817A}) identifying a similar two-component signature in both bursts is intriguing. It seems unlikely to be commonly detected in the SGRB population given the lack of identification prior to GRB 170817A despite thorough analyses of the gamma-ray data from SGRBs \citep{Spec_Catalog_BATSE_Preece_2000,Spec_Catalog_BATSE_Kaneko_2006,Three_Bright_SGRBs_Guiriec_2010,Spec_Catalog_Yu_2016, Spec_Catalog_KONUS_SGRB, Burns_Dissertation}. It is not clearly identifiable in two other close SGRBs, which is discussed in Section \ref{sec:tail}.
%, but this is not necessarily surprising. 
We report here a detailed analysis of GRB 150101B. We show it is a typical SGRB in gamma rays, and compare it with GRB 170817A. Despite their common morphology of a harder spike followed by a softer tail, the two bursts differ in important ways: GRB 150101B is neither subluminous nor subenergetic compared to other SGRBs, is not spectrally soft, and its total prompt duration in gamma rays is significantly shorter than GRB 170817A. Various theoretical models have been applied to explain the unusual behavior of the prompt and afterglow emission from GRB 170817A. We briefly comment on possible applicability of some of these models to GRB 150101B and discuss possible implications for GRB 170817A.

Prior to GW170817, \citet{GRB150101B_fong} presented an analysis of GRB 150101B, concluding that the afterglow exhibited typical broadband behavior for SGRBs. A more recent analysis of these data combined with a measure of the total gamma-ray energetics asserts that GRB 150101B has a luminous blue kilonova and an off-axis jet \citep{GRB150101B_Troja}. That analysis reports a measure of the total gamma-ray energetics that differs significantly from our previously published results \citep{GW170817-GRB170817A}. Their lower value arises from the selection of a narrower energy range and the differences between the \textit{Swift} Burst Alert Telescope (BAT) and \textit{Fermi} GBM observations of GRB 150101B, previously investigated in \citet{Burns_GBM_BAT}. A new understanding of the differing BAT and GBM observations of GRB 150101B is discussed in the Appendix. While investigating the differences between the GBM published values and those in \citet{GRB150101B_Troja} we discovered the characteristics of the burst presented here.

\section{Observed Properties of GRB 150101B}
GRB 150101B triggered GBM on-board at 2015-01-01 15:24:34.468 UTC, was reported to the community within 7 seconds, and cataloged with trigger ID GRB150101641 and trigger number 441818617\footnote{\url{https://gcn.gsfc.nasa.gov/other/441818617.fermi}}. A ground analysis of BAT slew data identified a significant source and constrained the position to (RA, Dec) =  (188.044, -10.956) with an uncertainty of 2.5 arcmin \citep{GCN_GRB150101B_BAT}, enabling broadband follow-up observations.

\begin{figure*}
\centering
\includegraphics[width=1.0\textwidth]{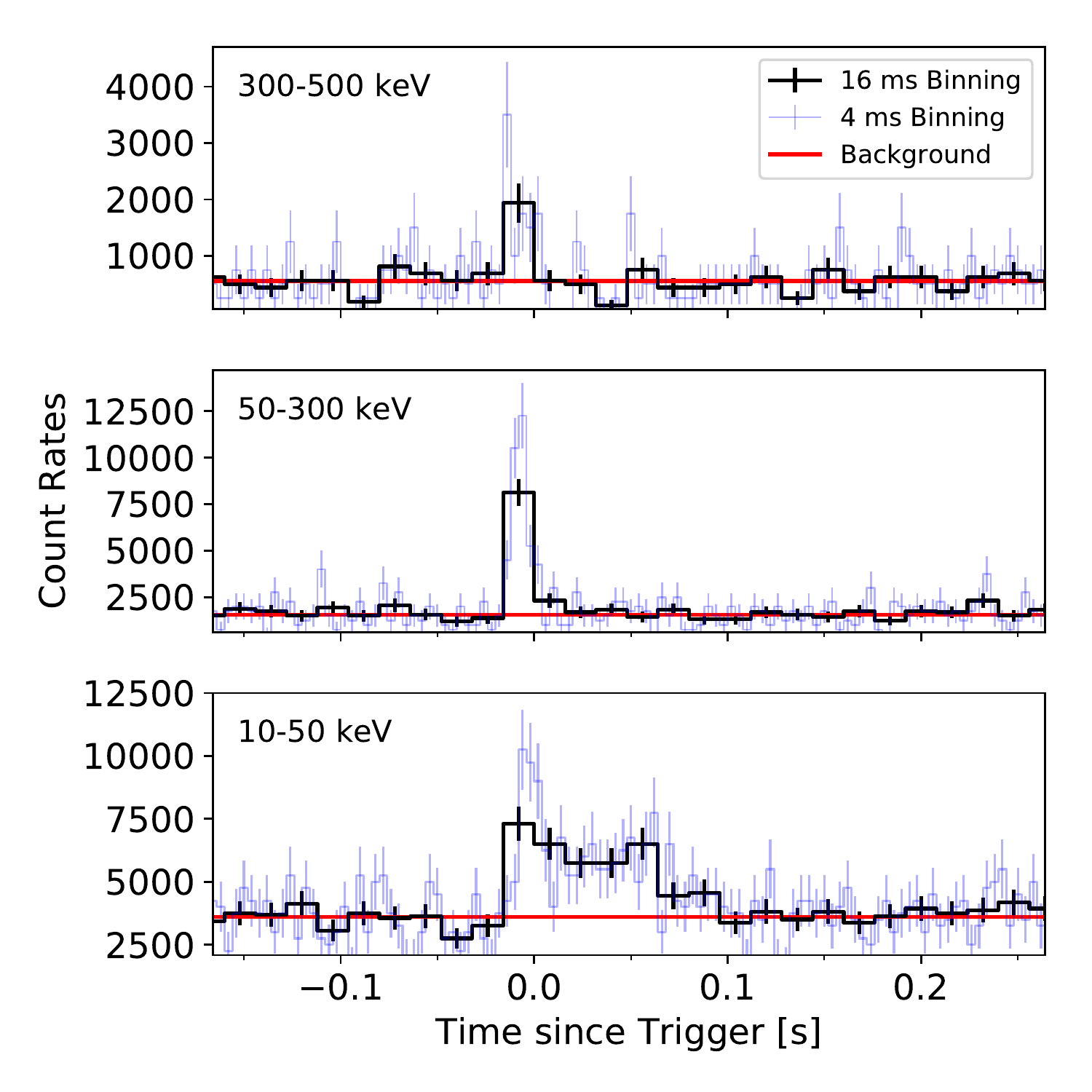}
\caption{The count rates in different energy ranges showing the short hard spike and the longer soft tail in GRB 150101B. All ranges include counts in the NaI detectors; the counts in the BGO detectors are included only in the highest energy range. The short hard spike is visible above 50 keV. The soft tail is obvious in the 10-50 keV channel. GRB 150101B triggered GBM on the 16 ms timescale corresponding to the main peak, which places T0 at the end of this interval. The background count rates around trigger time are flat and well behaved; the background levels shown here are the time-averaged values around the trigger.}
\label{fig:Counts}
\end{figure*}

The total duration, the two component nature, the durations of each component, and the significance of the soft tail in GRB 150101B are supported by several analysis methods. 
Figure \ref{fig:Counts} shows that the count rates recorded in the relevant GBM detectors as a function of time and energy are suggestive of two distinct spectral components. The counts are taken from all GBM detectors with good viewing geometry to the position of GRB 150101B at event time. These are the Sodium Iodide (NaI) detectors 3, 4, 6, 7, and 8 and both Bismuth Germanate (BGO) detectors. The reference time and combination of detectors is used throughout the analysis. The short hard spike is 16 ms long (covering the 16 ms before T0) followed by a longer soft tail that lasts about 64 ms (covering the 64 ms after T0), for an overall duration of about 80 ms. 

The standard GBM duration parameters, which describe the time to accumulate the central 90\% and 50\% of the total fluence, are T$_{90} = 0.08 \pm 0.93$ s and T$_{50} = 0.016 \pm 0.023$, respectively. These duration measures include contributions from both components. All GBM catalog values are taken from the online catalog\footnote{\url{https://heasarc.gsfc.nasa.gov/W3Browse/fermi/fermigbrst.html}}. Both measures start 16 ms before the trigger time. The large uncertainties are driven by the short timescales of the event and the difficulty in applying this method to SGRBs with modest fluence, but the start times and central values match those inferred from other methods. Few GBM SGRBs appear to have durations less than $\sim$100 ms long.

The timescales of each separate component are additionally supported by results from the GBM Targeted Search (discussed in the next section) and the Bayesian Block technique \citep{BB_scargle_2013}. Applying the latter technique to the data from the relevant GBM NaI detectors reveals the two emission episodes without any prior assumptions on timescales. In the 50-300 keV range, where GBM is most sensitive, the analysis isolates the short hard spike over the same pre-trigger 16 ms timescale as other methods. In the 10-50 keV energy range the analysis identifies the soft tail over a 73 ms interval starting at trigger time, with a significance of more than 10 sigma. This suggests the soft tail is marginally longer than 64 ms and is consistent with the slight excess in the succeeding 16 ms bins in the low-energy counts lightcurves (Figure \ref{fig:Counts}). To ensure both components arise from GRB 150101B, we localize them independently using the GBM Targeted Search method \citep{GBM_only_paper} and find that both are consistent with the known source position from \citet{GCN_GRB150101B_BAT}.

\begin{figure*}
	\subfigure{\label{fig:Liso}\includegraphics[width=0.5\textwidth]{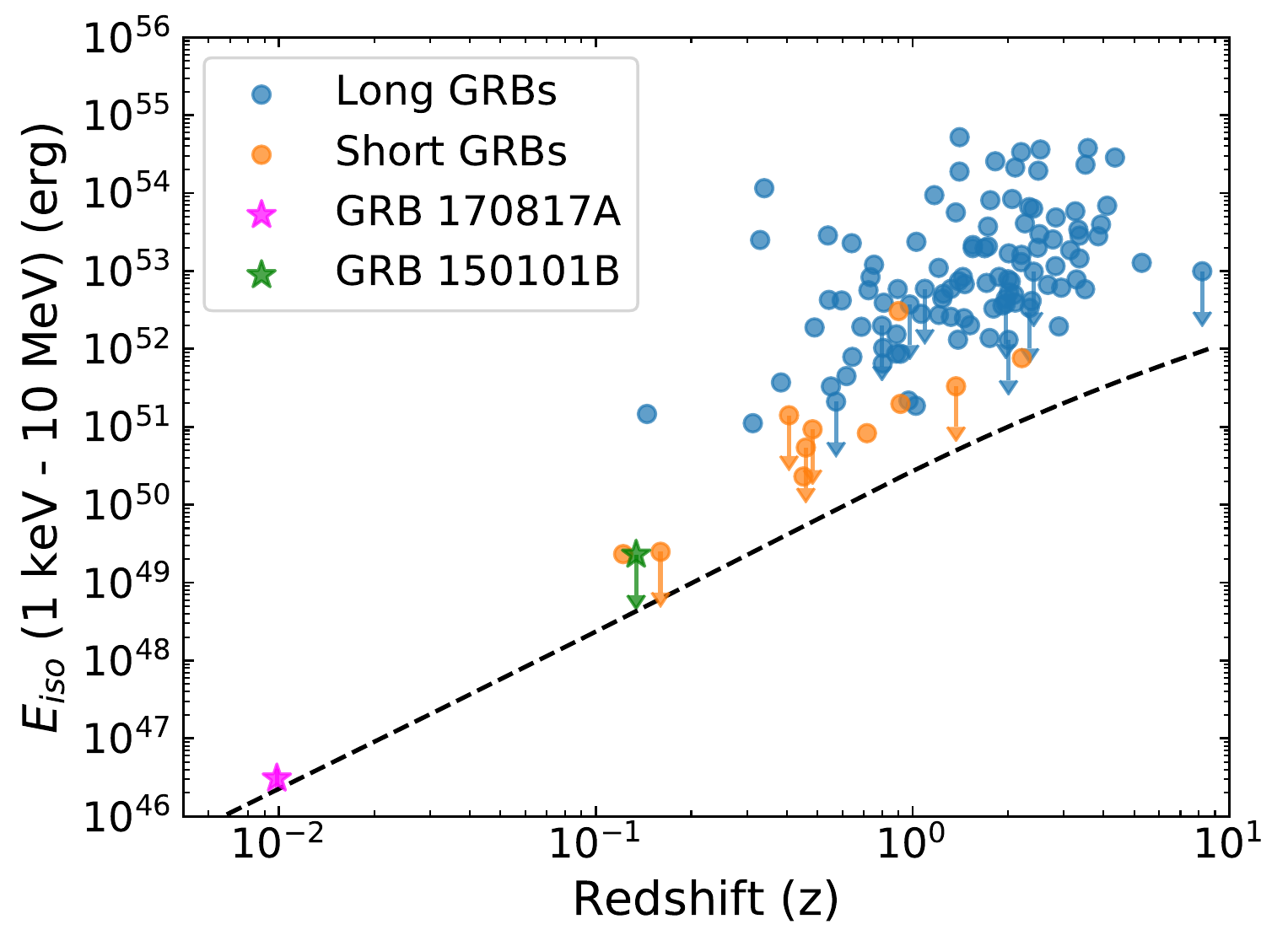}}
	\subfigure{\label{fig:Eiso}\includegraphics[width=0.5\textwidth]{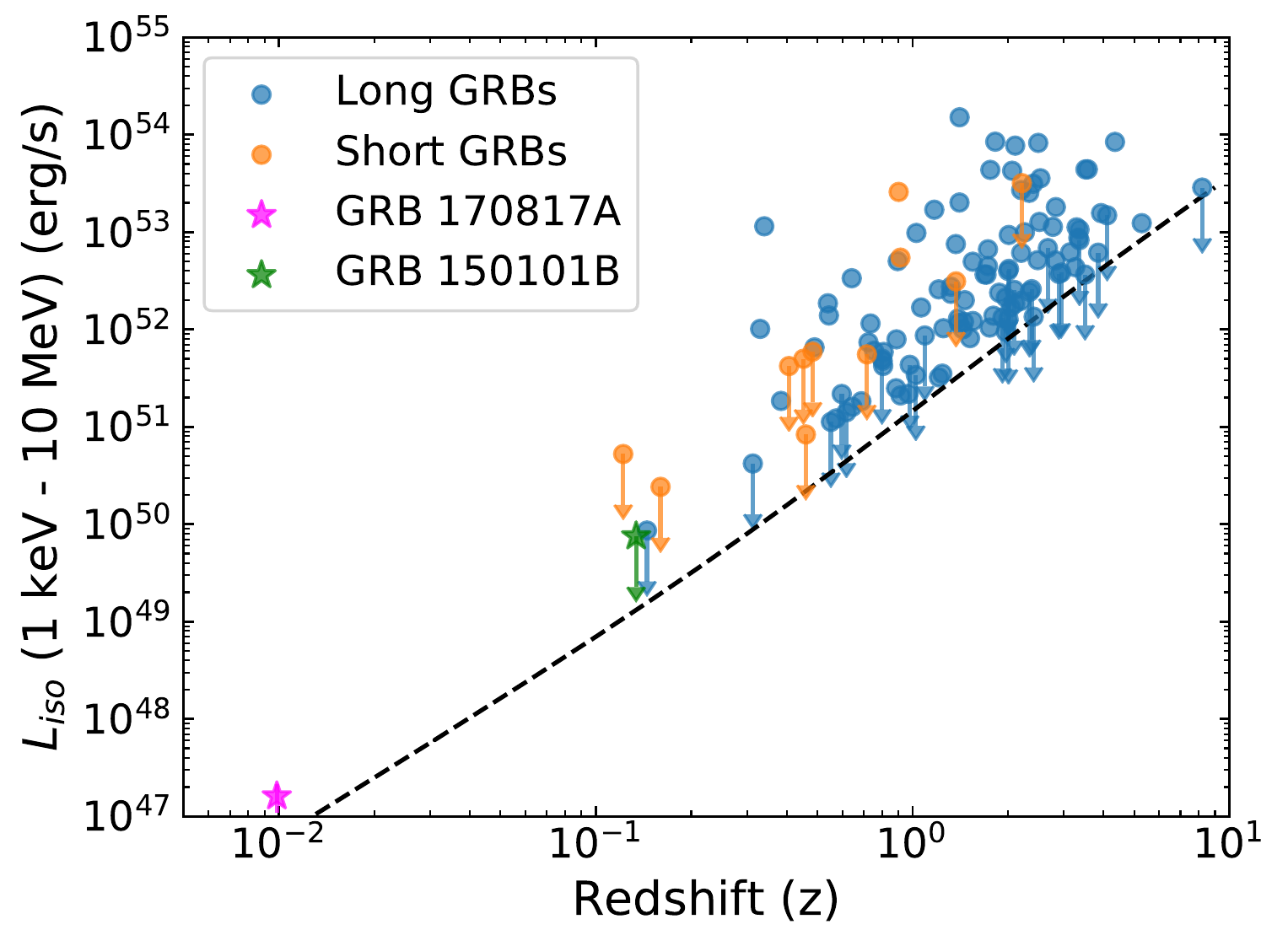}}
	\caption{{The isotropic equivalent energetics for GBM GRBs with redshifts. This is a modified version of a figure in \citet{GW170817-GRB170817A}. The total energetics (E$_{{\rm iso}}$) is shown on the left and the 64 ms peak luminosity (L$_{{\rm iso}}$) on the right. Both are given over the bolometric energy range from 1 keV to 10 MeV. GRBs best fit by a spectral model that constrains spectral curvature are shown as points. Power-law fits do not constrain spectral curvature and therefore overestimate the true value. GRBs best fit by a power-law are shown as upper limits. The dashed black line is an approximate threshold for triggering GBM on-board.}%{\it [Left]} 
\label{fig:isotropic_energetics}}
\end{figure*}

For GRBs with measured redshift and broad-band spectral observations we can constrain the total intrinsic energetics. By measuring the brightness at Earth and assuming constant flux over a sphere with a radius of the distance to the source we can measure isotropic-equivalent energetics, which are measured over the bolometric 1 keV - 10 MeV rest-frame energy range \citep{Iso_Energetics_Bloom_2001}. Some SGRBs have measured jet opening angles \citep{Rhoads_99,Fong15}. For collimated outflows isotropic-equivalent energetics are overestimates of the true intrinsic energetics. The two measures are the peak luminosity L$_{{\rm iso}}$, calculated from the peak flux, and the total energy E$_{{\rm iso}}$, calculated from the fluence. For GRB 150101B, the GBM catalog fluence is (2.4 $\pm$ 0.2)$\times$10$^{-7}$ erg/cm$^2$. This fluence is taken from the analysis that derives T$_{90}$, as it more fully captures the fluence of the burst by accounting for spectral evolution and contribution from low-count bins. The standard time-integrated and 64 ms peak flux and fit parameters from spectral analysis are given in Table \ref{tab:spectral_results}. All measures of flux and fluence are given over the 10-1000 keV energy range. The peak flux and fluence are converted to peak L$_{\rm iso}$ and E$_{{\rm iso}}$ assuming the redshift of the associated host galaxy z=0.134 \citep{GCN_GRB150101B_z,GRB150101B_fong}. The standard E$_{\rm iso}$ value for GRB 150101B is 2.3$\times$10$^{49}$ erg \citep{GW170817-GRB170817A}. Figure \ref{fig:isotropic_energetics} shows the E$_{{\rm iso}}$ and L$_{\rm iso}$ distribution for GRBs, which show two SGRBs with comparable total energetics. Also shown is GRB 170817A, which is a significant outlier in both distributions \citep{GW170817-GRB170817A}.

\begin{figure*}
	\subfigure{\label{fig:Energy_Flux}\includegraphics[width=0.5\textwidth]{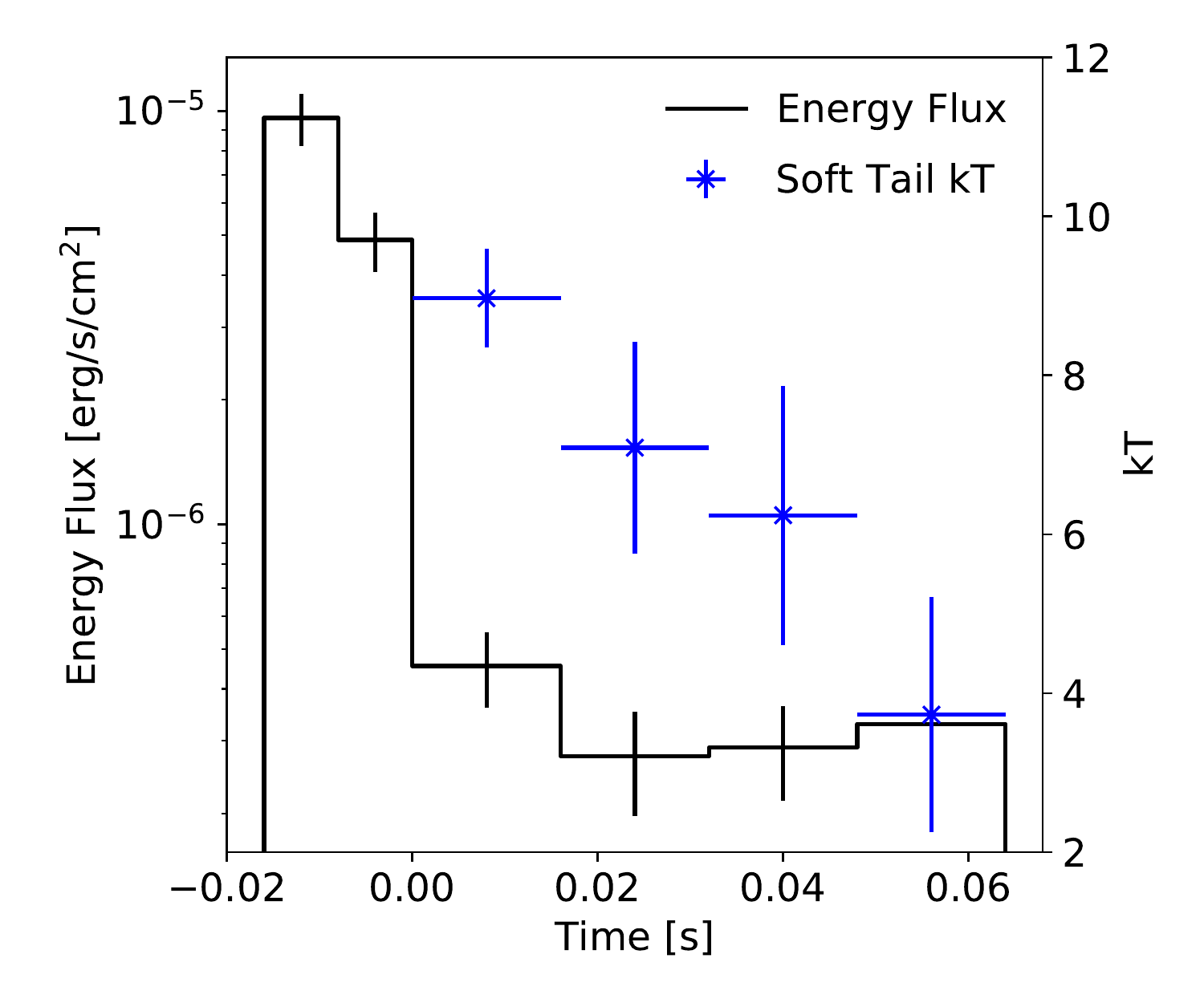}}
	\subfigure{\label{fig:Photon_Flux}\includegraphics[width=0.5\textwidth]{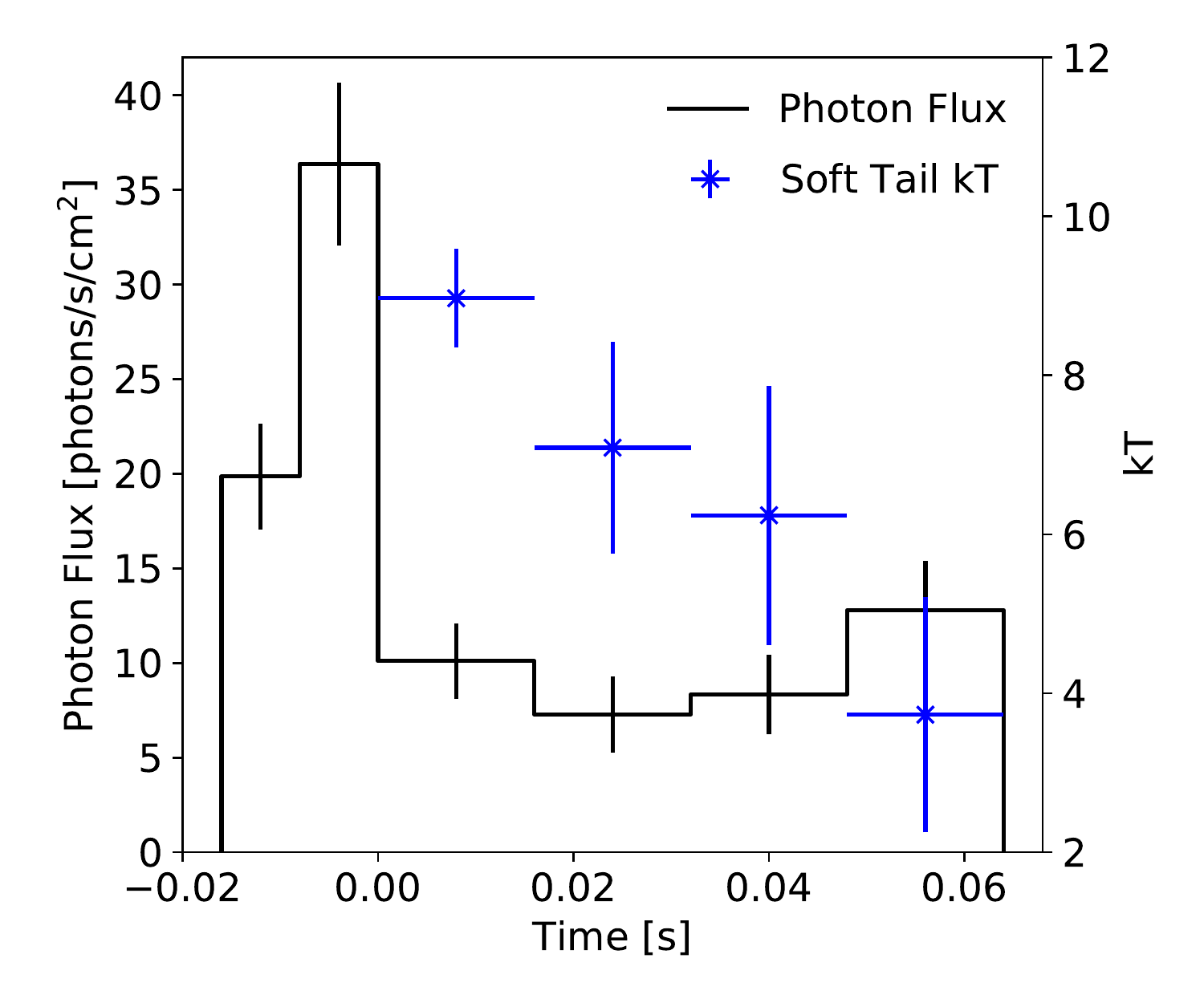}}
	\caption{{The deconvolved lightcurves of GRB 150101B in energy flux (left) and photon flux (right) shown in black. The main peak has significantly higher flux than the soft tail. The fits to the soft tail show a declining temperature, while the flux appears constant.}%{\it [Left]} 
\label{fig:GRB150101B_Flux}}
\end{figure*}

GRB 150101B has significantly more counts in GBM than GRB 170817A, which enables more detailed spectral analysis. To explore the best spectral fits to the two components in GRB 150101B and to search for evidence of spectral evolution, the burst is divided into time slices and data from each slice analyzed separately. 
Details of GBM spectral analysis and a description of the functions typically used in GRB spectral fits can be found in the GBM spectral catalogs \citep{GBM_catalog_Gruber}. The preferred models, best fit parameters, and fluxes from both the catalog results and the time-resolved fits are shown in Table \ref{tab:spectral_results} and Figure \ref{fig:GRB150101B_Flux}. The hard spike is best fit by a power law in energy with a flux that drops exponentially above a peak energy, referred to as a comptonized function. The peak energy lies close to the center of the distribution of peak energies for SGRBs detected by GBM. The soft tail is best fit by a blackbody spectrum; however, this does not mean the true spectrum is thermal. Using these best-fit functions, the fluence of the main peak is (1.2$ \pm$ 0.1)$\times$$10^{-7}$ erg/cm$^2$ and (2.0 $\pm$ 0.2)$\times$$10^{-8}$ erg/cm$^2$ for the soft tail. 

\begin{table*}
\centering
\begin{tabular}{| c | c | c | c | c | c | c |}% c |}
\hline
Time Range & Model & E$_{{\rm peak}}$ or kT & Index & Photon Flux & Energy Flux & L$_{{\rm iso}}$ \\%&  Fluence \\
(ms) & & (keV) & & ph/s/cm$^2$ & $10^{-7}$ erg/s/cm$^2$ & 10$^{49}$ erg/s\\%& $10^{-8}$ erg/cm$^2$ \\
\hline
\hline
Catalog & & & & & & \\
-64 : 64 & Power law & - & -1.8 $\pm$ 0.1 & 7.9 $\pm$ 0.9 & 8.3 $\pm$ 1.4 & -\\
0 : 64 & Power law & - & -2.4 $\pm$ 0.3 & 10.4 $\pm$ 1.4 & 4.8 $\pm$ 1.1 & $<$7.5 \\
\hline
\hline
Integrated & & & & & & \\%& \\
-16 : 0 & Comptonized & 550 $\pm$ 190 & -0.8 $\pm$ 0.2 & 28.3 $\pm$ 2.6 & 72 $\pm$ 8 & 44 $\pm$ 5 \\%& 13 $\pm$ 2 \\
0 : 64 & Blackbody & 6.0 $\pm$ 0.6 & - & 9.3 $\pm$ 1.1 & 3.1 $\pm$ 0.4 & 1.8 $\pm$ 0.2 \\%& 2.1 $\pm$ 0.3\\
\hline
\hline
Resolved & & & & & & \\%& \\
-16 : -8 & Comptonized & 1280 $\pm$ 590 & -0.4 $\pm$ 0.3 & 19.8 $\pm$ 2.8 & 96 $\pm$ 14 & 101 $\pm$ 15\\%& - \\
-8 : 0 & Comptonized & 190 $\pm$ 50 & -0.7 $\pm$ 0.3 & 36.4 $\pm$ 4.3 & 49 $\pm$ 8 & 26 $\pm$ 4 \\%& - \\
				
0: 16 & Blackbody & 9.0 $\pm$ 1.3 & - & 10.1 $\pm$ 2.0 & 4.5 $\pm$ 1.0 & 2.4 $\pm$ 0.5 \\
16 : 32 & Blackbody & 7.1 $\pm$ 1.6 & - & 7.3 $\pm$ 2.0 & 2.7 $\pm$ 0.8 & 1.5 $\pm$ 0.5 \\
32 : 48 & Blackbody & 6.2 $\pm$ 1.5 & - & 8.4 $\pm$ 2.1 & 2.9 $\pm$ 0.7 & 1.7 $\pm$ 0.4 \\
48 : 64 & Blackbody & 3.7 $\pm$ 0.7 & - & 12.8 $\pm$ 2.6 & 3.3 $\pm$ 0.7 & 2.5 $\pm$ 0.5 \\
\hline
\end{tabular}
\caption{The preferred models and best fit parameters from the spectral analysis of GRB 150101B. For the Catalog rows the -64 to +64 ms interval corresponds to the fluence measure from the GBM spectral catalog, and the 0 to +64 ms interval to the peak flux interval. 
We compared the standard GRB functions, a blackbody, and multi-component fits of a blackbody and a standard GRB function. Single component fits are preferred in all intervals. The best fit models were either a power law, a blackbody, or a comptonized spectrum; see \citet{GBM_catalog_Gruber} for details. E$_{{\rm peak}}$ parameterizes the peak energy for the comptonized spectrum. The L$_{{\rm iso}}$ measures for intervals less than 64 ms cannot be directly compared to other bursts.}
\label{tab:spectral_results}
\end{table*}

From summing the time-resolved fits, the total energetics values for the main peak (MP) and soft tail (ST) are E$_{{\rm iso}}^{MP}$ = (9.0 $\pm$ 1.1)$\times$$10^{48}$ erg and E$_{{\rm iso}}^{ST}$ = (1.1 $\pm$ 0.1)$\times$$10^{48}$ erg. The sum of the E$_{{\rm iso}}$ of the two components is about half the E$_{{\rm iso}}$ upper limit inferred from the standard analysis, confirming the standard analysis as a reliable measure. The L$_{{\rm iso}}$ values of each component, as well as sub-intervals, are given in Table \ref{tab:spectral_results}. The very short duration of the main peak and the two component composition complicate the standard calculation of L$_{{\rm iso}}$ for GRB 150101B. Nevertheless, L$_{{\rm iso}}^{MP}$ $\approx$ 4$\times$10$^{50}$ erg/s, within the normal distribution of SGRBs, as evident from Figure \ref{fig:isotropic_energetics}.

As shown in Table \ref{tab:spectral_results} and Figure \ref{fig:GRB150101B_Flux} we further resolve each component: the main peak into two 8 ms intervals and the soft tail into four 16 ms intervals. We caution against strong inferences from fits to such short timescales; we use them here to investigate possible spectral evolution. The time-resolved fits of the main peak suggest hard to soft evolution within the pulse, seen in most GRBs \citep{Spec_Catalog_Yu_2016}. The time-resolved fits of the soft tail show a declining temperature, while the flux is approximately constant.

%For both 8 ms bins of the main peak power law with an exponential cutoff is preferred. The first bin has a peak energy of $\sim$1 MeV, which is consistent with the high counts in the BGO detectors. The second 8 ms bin is significantly softer. In all four 16 ms bins of the soft tail the blackbody spectrum is slightly preferred over a simple power law. The soft tail is consistent with constant flux and decaying temperature, providing marginal evidence for cooling \textcolor{red}{Peter V was looking into quantifying this. Results?}.

The minimum variability timescale for GRB 150101B, measured in the same manner as \citet{Min_Variability_Golkhou_2015}, is 7.5 $\pm$ 0.8 ms. This millisecond variability constrains the ratio of the radius (R) of the emitting region to the bulk Lorentz factor ($\Gamma$) to a value R / $\Gamma^2$ $<2 c t_v/(1+z)\approx$ 4000 km. With $\Gamma \approx$1200 (the highest lower limit inferred for a SGRB, from \citealt{090510_Fermi_2010}) we constrain R $<$ 5.7$\times$10$^{9}~(\Gamma/1200)^2$ km.

\section{Comparison to GRB 170817A}
\begin{figure*}
\centering
\includegraphics[width=1\textwidth]{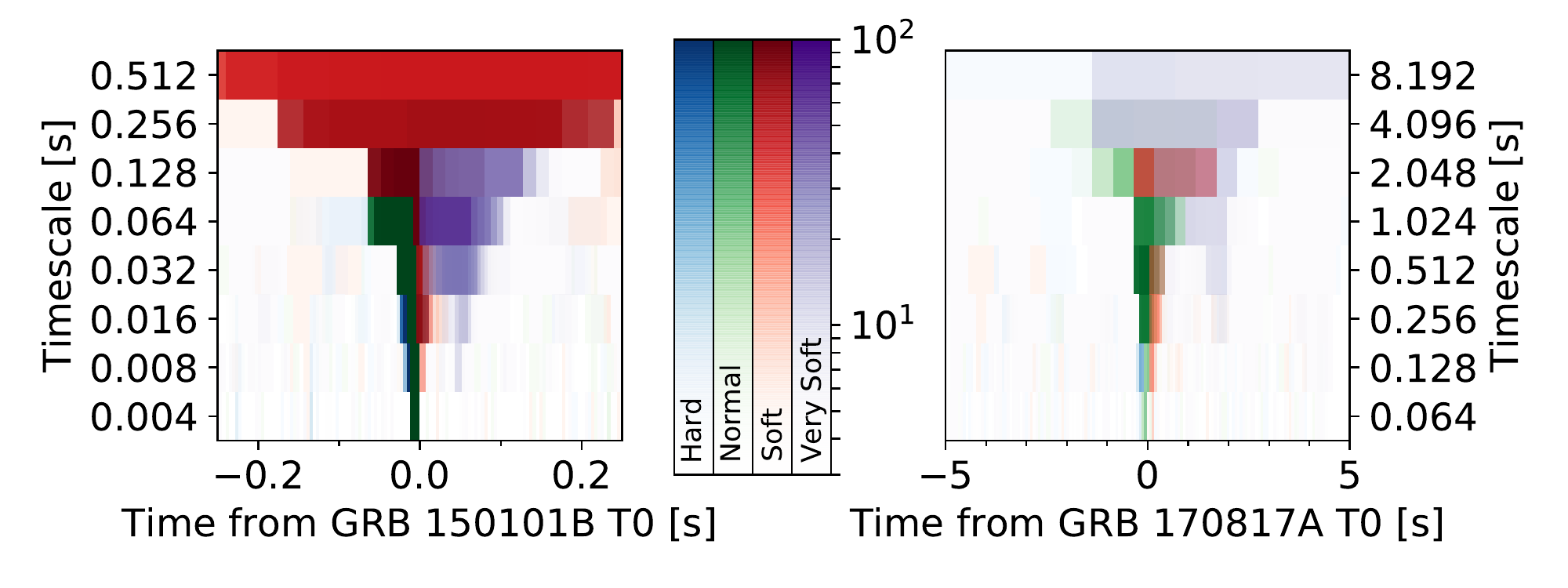}
	\caption{{The similarity of GRBs 150101B (left) and 170817A (right) in a single image: the spectrally separated waterfall plots from the GBM Targeted Search runs. Times are relative to the trigger times of the respective bursts. The four templates used here include three GRB-like spectra (the hard shown in blue, normal in green, and soft in red) and one very soft thermal template (shown in purple; kT = 10 keV). The color maps show the log likelihood ratio for that template; all are fixed to the same range. This last template was added in response to the discovery of the soft tail in GRB 170817A. The bursts are phenomenologically similar.}%{\it [Left]} 
\label{fig:waterfall_spec}}
\end{figure*}

Figure \ref{fig:waterfall_spec} shows the key features in GRBs 150101B and 170817A: a main peak characterized as a short hard spike followed by a transition to longer softer emission with a possible secondary peak, that is reasonably consistent with a thermal spectrum. 
This figure is generated with the GBM Targeted Search \citep{Blackburn2015,Goldstein2016,Targeted_Search_Kocevski_2018} and shows as a function of time the characteristics of the most significant signals above background revealed by the search on each of the input timescales. 
The color coding reflects the spectral template favored by the transient and the intensity maps the log likelihood ratio that a source is preferred over just background in that time interval. 

For GRB 150101B the search identifies the soft tail most significantly on the 64 ms timescale starting at T0. The signal in this source window has a log likelihood ratio of about 60 for the very soft thermal spectral template, which is more significant than several GBM-triggered SGRBs \citep{Targeted_Search_Kocevski_2018}. The hard spike is found with maximal significance in the 16 ms source window ending at trigger time, with a log likelihood ratio greater than 400. These source windows of maximum significance match the durations and phases previously inferred by visual inspection of the counts lightcurves and the Bayesian Blocks analysis, providing additional evidence for the two components. The transition from the main peak to the thermal-like tail is stark for GRB 150101B, occurring immediately once the 64 ms source window no longer overlaps with the main peak. The fast transition and the soft tail persisting for several times the duration of the main peak is unlike typical hard to soft evolution observed in GRB pulses \citep{Spec_Catalog_Yu_2016}. 

However, there are important differences between GRBs 150101B and 170817A. GRB 170817A has intrinsic isotropic energetics several orders of magnitude below any other SGRB with known redshift; GRB 150101B does not. The ratios of the peak luminosities of the main peak to the soft tail in GRB 150101B are far greater than the factor of a few difference for GRB 170817A \citep{GW170817-GRB170817A}. The main peak of GRB 150101B appears to have a higher peak energy than that of GRB 170817A, but they are roughly consistent within errors (GRB 170817A also shows spectrally harder intervals in time-resolved analysis, see \citealt{Veres_GRB170817A_2018}). The greatest difference between the two bursts inferred from GBM data alone is the absolute timescale: GRB 150101B has an observed duration more than an order of magnitude shorter than GRB 170817A.

\section{The Origin of the Soft Tail}
Historically GRBs have been modeled as uniform top-hat jets because they sufficiently explained observations. When jets plow through dense environments they deposit a fraction of their energy in a hot cocoon %produce cocoon emission 
\citep{cocoon_ramirezruiz, cocoon_Peer_06}, which may occur in binary neutron star mergers as the ultrarelativistic jet that powers the SGRB pushes through material ejected just before merger \citep{cocoon_lazzati_2016}. GRB 170817A has odd behavior in both prompt and afterglow emission; the origin of which is a matter of some debate \citep{GW170817-GRB170817A,GW170817_xray_Troja,170817_Margutti_1,170817_Alexander_Xray_decline,GW170817_Haggard_1_2017,170817_Kasliwal,170817_bromberg_2017,170817_Mooley_1_2018,170817_nakar_2018,Veres_GRB170817A_2018,170817_troja_2_2018,170817_lazzati_2018,170817_gottlieb_2018,GW170817_offaxis_Alexander_2017,170817_Lyman_2018,170817_nynka_2018,170817_Ruan_2018,170817_VLBI}. Possible interpretations include a structured ultrarelativistic jet (e.g. \citealt{170817_Alexander_Xray_decline}), a jet and cocoon together (e.g. \citealt{GW170817-GRB170817A}), or a cocoon model (e.g. \citealt{170817_Kasliwal}) where the shock breakout produces the harder peak. Possible mechanisms for the soft tail include the photosphere of the jet or arising from the photosphere of the cocoon. % emission.

Some argue that both the main peak and soft tail of GRB 170817A can be explained by cocoon shock breakout-models \citep{170817_Kasliwal,cocoon_gottlieb}. For GRB 150101B L$_{{\rm iso}}^{MP} \sim$4$\times10^{50}$ erg/s, the L$_{{\rm iso}}$ ratio is $\sim$25, and the duration is an order of magnitude shorter than 0.5 s, all of which appear inconsistent with the simulations in \citet{cocoon_gottlieb}.
The properties of the shock breakout emission are determined by the radius where the breakout occurs and the shock velocity. From this, it follows that there should be a relation between observables such as duration, total energy and typical energy
\citep{Nakar+12shockbo}:
$T_{90} \approx 1~\rm s ~(E/10^{49} {\rm erg})^{1/2} (E_{\rm peak}/550~{\rm keV})^{-2.68}$. For GRB 150101B, the duration $<0.1$ s, which is incompatible with the cocoon shock breakout model. The soft tail with typical energy of $3.9\times$ 6.0 kT$\sim$23 keV and similar duration is even more inconsistent with the above relation. This conclusively excludes cocoon shock breakout as the origin of the main peak of GRB 150101B. %However, the soft tail may arise from the wide-angle cocoon emission during the brightest phase.
However, the soft tail may arise from the photosphere of the wide-angle cocoon during the brightest phase.

%{\bf I am learning as we go here: Nakar Piran et al say that the whole emission is from when a shock breaks out of the cocoon, so they specify the emission mechanism. On the other hand, there is the cocoon photosphere emission, where the cocoon is just like a jet, only dirtier and slower, and it produces the thermal radiation when it expands to its photosphere. That's why I added the correction in the last paragraph... }

Afterglow observations of GRB 170817A 200-300 days post-merger show a turnover in the temporal decay from X-ray to radio that appears to favor the structured jet scenario over the cocoon scenario \citep{170817_Alexander_Xray_decline}. The VLBI measurements of the movement of the core of the radio emission do as well \citep{170817_VLBI}. If the components in GRBs 150101B and 170817A are indeed similar, our analysis of the main peak of GRB 150101B adds additional evidence against the observed non-thermal emission originating entirely from a cocoon. 

In GRB 150101B, the onset of the soft tail emission begins at least by the end of the main peak, but may occur earlier and be hidden by the main peak. This suggests the significantly more energetic main peak and the soft emission have a common origin. While the absolute timescales of the two bursts differ greatly, the relative timescales are similar: the duration of the soft tail is about four times as long as the duration of the main peak and the possible secondary, very soft peak arises at the end of this interval. %With only two examples this could arise from chance but, if not, it is suggestive of a relation between the two components. 
The long and softer observed characteristics of GRB 170817A could be a result of timescale broadening, Dt $\propto$ 1 + ($\Gamma$ $\theta_{{\rm off-axis}}$)$^2$ \citep{GW170817-GRB170817A}. This is consistent with the inclination angle inferred from multimessenger observations \citep{170817_inclination_Finstad_2018,170817_LIGO_updated_parameters} and follow-up observations of the non-thermal emission \citep{170817_Margutti_1,170817_VLBI}. The short minimum variability timescale for GRB 150101B suggests a high $\Gamma$ \citep{Sonbas+15LF} and the higher luminosity would be expected for a more on-axis or fully on-axis alignment. If both components emit from the structured jet then they would be significantly broadened for GRB 170817A but not for GRB 150101B, resulting in large differences in absolute timescales but not in the relative timescales of the two components. Therefore, this interpretation is self-consistent for both bursts. One physical model for both components arising from the jet is an external shock origin for the main peak and a photospheric origin for the soft tail \citep{GW170817-GRB170817A}. Subdominant spectral components that are consistent with a thermal origin are often observed during the main peak of GRBs and generally attributed to photospheric emission \citep{ryde05}. The soft tail would follow the main peak if the photospheric radius is larger than the deceleration radius \citep{GW170817-GRB170817A}.

We here demonstrate that this model can explain GRB 150101B. The innermost stable circular orbit for a 2.8M$_{\odot}$ black hole (roughly the total mass of GW170817; \citealt{GW170817-GW}) is R$_0$=2.5$\times10^6$ cm. We can rewrite this as R$_0=10^{6.4}~$R$_{0,6.4}$ cm, a notation we use in the following derivation. A jet launched at R$_{0,6.4}$ with a total luminosity L$_0=10^{51}~$L$_{0,51}$ erg/s will have an initial temperature of kT$_0$=k(L$_0$/4$\pi$R$_0^2$ca)$^{1/4}$=1.3~L$_{0,51}^{1/4}$R$_{0,6.4}^{-1/2}$ MeV, with the radiation constant a=7.57$\times10^{-18}$ kg/s/cm$^2$/K$^4$. The jet accelerates as $\Gamma\propto$R/R$_0$ until it reaches the saturation radius R$_{\rm sat}=\eta$R$_0=7.2\times10^8\eta_{2.5}$R$_{0,6.4}$ cm, with $\eta$ the dimensionless entropy of the fireball (which is generally above $\sim$100 to prevent the compactness problem; \citealt{Goodman1986ApJ}). 

The jet becomes optically thin at the photospheric radius R$_{\rm phot}$=L$\sigma_T$/4$\pi$m$_p$c$^3\eta^3$=4.3$\times10^{10}$~L$_{0,51}\eta_{2.5}^{-3}$ cm (when neglecting pairs; \citealt{Meszaros+00phot}), with $\sigma_T$ the Thompson cross section and m$_p$ the mass of the proton. Here the photosphere occurs during the coasting phase (R$_{\rm sat}<$R$_{\rm phot}$), giving an observed temperature kT$_{\rm obs}$=kT$_0$(R$_{\rm phot}$/R$_{\rm sat})^{-2/3}$=3.5~L$_{0,51}^{-5/12}\eta_2^{8/3}$R$_{0,6.4}^{1/6}$ keV and L$_{\rm phot}$=L$_0$(R$_{\rm phot}$/R$_{\rm sat})^{-2/3}$=2.7$\times10^{48}~$L$_{0,51}^{1/3}\eta_2^{8/3}$R$_{0,6.4}^{2/3}$ erg/s. For fiducial values L$_0\approx{\mathbf {6.1}}\times10^{\mathbf {51}}$ R$_{0,6.4}^{-2/3}~$erg/s (L$_0$ exceeds L$_{iso}$ because it also converts into the kinetic energy of the jet) and $\eta\approx160$ R$_{0,6.4}^{-1/6}$ in the above equations, we recover the observed kT$_{\rm obs}=6$ keV and L$_{\rm phot}$=1.8$\times 10^{49}$ erg/s.

Lastly, we derive the condition for the photospheric radius to occur above the deceleration radius. The density in the close vicinity of a neutron star merger can be approximated as a wind medium \citep[e.g.][]{Bauswein+13bns}.
The number density can be written as  $n(R)$=AR$^{-2}$ where A=$\dot{M}/4\pi$m$_pv$, with $v$ the outflow velocity of the wind and $\dot{M}$ is the mass-loss rate. The deceleration radius, where the jet slows down significantly is:
R$_{\rm dec}$=E$_k$/4$\pi$m$_p$c$^2$A$\eta^2$ 
where E$_k$ is the kinetic energy of the outflow \citep{panaitescu00}. For R$_{\rm dec}\lesssim$R$_{\rm phot}$, A$\gtrsim4.5\times10^{35}\eta_2$E$_{k,50}$L$_{0,51}^{-1}$. This corresponds to a mass density $\rho\gtrsim10^{-2}~ ($R$/100~{\rm km})^{-2}{\rm g/cm}^{3}$, which is consistent with simulations \citep{Bauswein+13bns}. Therefore, the condition for the soft tail arising from photospheric emission and being emitted at a larger radius than the external shocks that produce the main peak matches expectations for BNS mergers. We conclude this model can explain GRB 150101B, as well as GRB 170817A \citep[as shown in][]{GW170817-GRB170817A}.

\section{The Nature and Detectability of the Soft Tail}
\label{sec:tail}
While the significance of the soft tail in GRB 150101B is unambiguous ($>$10 sigma, according to the Bayesian Blocks analysis), its origin is uncertain. There is evidence in favor of the soft tail originating from thermal emission: there is a statistical preference for a blackbody spectrum (over standard GRB functions) in both the time-integrated and time-resolved fits of the soft tail, the parameters of a comptonized spectrum being similar to those that would arise from a thermal component (high index, matching peak energies), and evidence of cooling. Together this suggests the tail may have a thermal origin but is not conclusive. The forward-folding technique can only measure the consistency of the data with an assumed function and other emission mechanisms or multi-temperature emission zones may result in blackbody-like emission and evolution to softer values. Therefore, just as for GRB 170817A, we cannot be certain the soft tail is thermal in nature \citep{GBM_only_paper}.  

% Under the assumption that the soft tail is thermal and quasi-spherical, L$_{iso}$ = 4$\pi$R$^2$$\sigma$T$^4$, where $\sigma$ is the Stefan-Boltzmann constant. For the time-integrated fit of the soft tail we can infer a radius of the emitting region R$_{thermal}$ = (3.1 $\pm$ 1.1) $\times$ 10$^5$ km. \textcolor{red}{Maybe remove this section? It isn't used anywhere.} %This compares to our upper limit on the radius of the jet, R$_{jet}$ $<$ 5.7 $\times$ 10$^{9}~(\Gamma/1200)^2$ km. %R$_{thermal}$ is $\sim$10\% of our upper limit on R$_{jet}$, suggesting the thermal emission would be mildly relativistic. This matches the inferred velocities for possible cocoon emission in GRB 170817A \textcolor{red}{cite}. %\textcolor{red}{measure R as a function of time with time-resolved fits. Can we measure v/c?}

Figure \ref{fig:GRB150101B_Flux} and Table \ref{tab:spectral_results} show the evidence for cooling in the soft tail. GBM can constrain blackbody temperatures down to kT$\sim$3 keV \citep{xray_burst_catalog_jenke}, the last of the four 16 ms bins of the soft tail has a temperature of about 4 keV, and the next two 16 ms bins are $\sim$2 sigma above background in the lowest energy range. All of this is consistent with a blackbody at $\sim$10 keV just after T0, which cools out of detectability in less than 100 ms. If the soft tail is not thermal this statement likely still holds, as a blackbody spectrum reasonably captures its behavior. If true, we detect the soft emission for GRB 150101B only because the main emission is extremely short. More generally, soft emission would be unidentifiable in SGRBs if it cools out of detectability before the dominant non-thermal emission ends. However, this would also require the un-broadened tail of GRB 170817A to be detectable longer than the main emission episode.

Ground-based gravitational wave interferometers quantify their sensitivity by the detection range of canonical BNS (1.4M$_{\odot}$) mergers \citep[see, e.g.][]{Gen_2_Prospects_2018}. GW interferometers have position-dependent sensitivities; the range is the radius of the spherical equivalent volume that a given interferometer is sensitive to. The updated BNS merger detection range for Advanced LIGO is 173 Mpc \citep{LIGO_173Mpc}. The most distant events that can be detected are face-on mergers at the position of the maximal antenna pattern sensitivity, which is 2.26 times the detection range \citep[when cosmological effects can be neglected;][]{GW_horizon_Finn_1993}. Joint GW-GRB detections can extend this range by $\sim$20-25\% \citep{GW_GRB_williamson_2014,Blackburn2015,LVC_GW_GRB_joint_2017}. Altogether, the joint GW-GRB detection horizon of Advanced LIGO at design sensitivity is $\sim$500 Mpc \citep[z$\approx$0.1;][]{Burns_Dissertation}. GRB 150101B occurred at a redshift of 0.134 \citep{GCN_GRB150101B_z}, corresponding to a luminosity distance of $\sim$650 Mpc (using standard cosmology from \citealt{Planck15}), somewhat beyond where Advanced LIGO could detect a BNS merger. Using a nominal SNR threshold of $\sim$5.4 \citep{GBM_only_paper}, the GBM Targeted Search could recover the main peak to $\sim$1500 Mpc, and the soft tail to $\sim$900 Mpc. With the most sensitive search for short gamma-ray transients, the main peak of GRB 150101B has a detectable volume five times as large as the soft tail.

Two other close SGRBs are GRB 080905A and GRB 160821B. GRB 080950A has an associated host galaxy at z=0.1218 \citep{GRB080905A_redshift_Rowlinson_2010}. GRB 160821B has an associated host galaxy at z=0.16 \citep{GCN_GRB160821B_z}. While both have $>$10 kpc offsets from the center of their host galaxies, these intrinsic offsets are within the offset distribution for SGRBs (see \citealt{Fong15} for a compiled sample) and both putative host galaxies are large (giving reasonable offsets when normalized by the light radius of the host galaxy). From both visual inspection of the lightcurves, runs of the Targeted Search, and time-resolved spectral analysis, neither GRB 080905A nor GRB 160821B have obviously distinct tails. For reasons previously discussed, this is not necessarily surprising. Both SGRBs are about a second long and the soft tails may cool out of band before the main emission ends. Additionally, because GBM is a background-dominated instrument at low energies, shorter transients are more easily distinguished from background. Therefore it is more difficult to distinguish comparatively weak soft emission over longer timescales that would be expected for these two bursts if the relative timescales hold. The soft emission may also be undetectable in these bursts given the relative L$_{{\rm iso}}^{ST}$ values compared to L$_{{\rm iso}}^{MP}$ for GRBs 150101B and 170817A, i.e. they occur at a distance where the main peak is detectable but the soft tail is not recoverable. If the soft component is of cocoon origin, then the luminosity of the soft tail may depend on the uncertain ejecta density or the external density into which the jet and cocoon propagate. These densities may vary considerably between SGRBs, resulting in a wide range of luminosities for the secondary soft tails. It may also be that these bursts just do not have soft tails.

% \textbf{Evidence for cooling? Does it extend into the main peak?}

% \textcolor{blue}{Dan, Peter, etc, I'm happy for others to write theory/possible interpretation here.}

\section{Conclusion}
GRB 170817A was certainly a unique burst: the second multimessenger astrophysical transient (after SN1987A, see \citealt{SN1987A}), long and soft for a SGRB, subluminous, the first with a distance measured by gravitational waves, the closest SGRB with a known distance, and the apparent two-component nature. GRB 150101B is short and hard, has unexceptional properties detected at Earth, is one of the closest SGRBs, and has the same apparent two-component nature of the prompt gamma-rays as GRB 170817A. \citet{GRB150101B_Troja} argue that the follow-up observations of GRB 150101B show similarities to GRB 170817A.

Finding an unusual observational signature in one transient is extremely interesting. Finding that same two-component signature in prompt gamma-rays in a second nearby SGRB suggests it may be a common feature. If the soft tail is an intrinsic property of SGRBs it may have been previously unrecognized due to the lack of detectability (because it is subdominant to the main emission or the distance to the source too great) or may lie hidden in the data owing to insufficiently targeted analysis techniques. It may be more difficult, or even impossible, to detect in instruments less sensitive or with a higher low-energy threshold than GBM. 

SGRBs with extended emission \citep[see, e.g.][]{extended_emission_2001_lazzati,extended_emission_2002_connaughton,extended_emission_2006_norris} are SGRBs with a usual short hard spike and fainter emission lasting for tens to $\sim$100 s, where the extended emission may have a higher fluence than the short spike. It appears unlikely that this emission is similar to the soft thermal-like tails observed in GRBs 150101B and 170817A. Extended emission tends to be softer than the hard spike, but some have peak energies of several hundred keV \citep[e.g.][]{extended_emission_kaneko_2015,extended_emission_svinkin_2016}; in some cases the peak energies of the extended emission exceeds that of the main pulse \citep{extended_emission_svinkin_2016}. Some fits of extended emission can constrain spectral curvature and give low-energy power-law indices similar to those observed for GRBs \citep{extended_emission_kaneko_2015,extended_emission_svinkin_2016}, which is dissimilar to the values expected for comptonized fits of a blackbody spectrum. No previously identified SGRB with extended emission appears similar to GRBs 150101B or 170817A, but these investigations predate GRB 170817A. A careful examination of the \textit{Fermi} GBM SGRB population will provide insight into the commonality and origin of the soft tail. The search of the full GBM SGRB population is the subject of an on-going study, and will be informed by the results of this work.

While SGRBs have been observed for 50 years, neutron star mergers are now studied in new ways as both gravitational waves and kilonovae. Gravitational wave observations provide a distance measure free from the ambiguity of associating SGRBs with their host galaxies in the nearby universe \citep{Tunnicliffe_hostless}, a time and location to seed the GBM Targeted Search (enabling the detection of GRB 170817A-like events to greater distances, see \citealt{GBM_only_paper}), and provide direct observations of the central engine unavailable to electromagnetic observatories. Further, while GRB observations can suggest one of the compact object progenitors is a neutron star, gravitational wave or kilonova observations may be able to distinguish between binary neutron star and neutron star black hole mergers. Cocoon emission may always arise for binary neutron star mergers but is not expected to occur for neutron star-black hole mergers as they likely have lower densities at their polar regions \citep{kilonovae_LRR_metzger}, where the jet is believed to originate. Therefore, future multimessenger observations may determine if the soft tail arises from a cocoon. 

However, if the soft tail is confirmed with future observations or more careful analysis of existing observations, and is generally subdominant, it will be a key observational signature to identify nearby events from the prompt gamma-ray emission alone, regardless of the physical mechanism. This would enable the prioritization of follow-up gravitational wave searches or electromagnetic follow-up observations shortly after the time of the merger, and provide unique insights into the physics of neutron star mergers.

\acknowledgments
The authors thank the referee for their valuable comments and feedback. The authors thank A. Lien for an exhaustive effort to help us understand the differences between the GBM and BAT observations of GRB 150101B.
The authors also thank M. Aloy for the combined external shocks and photospheric emission initially presented in \citet{GW170817-GRB170817A}.
The USRA co-authors gratefully acknowledge NASA funding through contract NNM13AA43C. 
The UAH co-authors gratefully acknowledge NASA funding from co-operative agreement NNM11AA01A. 
E.B. and T.D.C. are supported by an appointment to the NASA Postdoctoral Program at the Goddard Space Flight Center, administered by Universities Space Research Association under contract with NASA. 
D.K., C.A.W.H., C.M.H., and T.L. gratefully acknowledge NASA funding through the \textit{Fermi} GBM project. 
Support for the German contribution to GBM was provided by the Bundesministerium f{\"u}r Bildung und Forschung (BMBF) via the Deutsches Zentrum f{\"u}r Luft und Raumfahrt (DLR) under contract number 50 QV 0301. 
A.v.K. was supported by the Bundesministeriums für Wirtschaft und Technologie (BMWi) through DLR grant 50 OG 1101. 
N.C. acknowledges support from NSF under grant PHY-1505373.

\bibliographystyle{aasjournal}
\bibliography{references.bib}

\appendix
\section{Other Prompt Gamma-ray Observations of GRB 150101B}
GRB 150101B triggered \textit{Fermi} GBM on-board \citep{GCN_GRB150101B_GBM} and was found in ground searches of the data from the \textit{Swift} BAT \citep{GCN_GRB150101B_BAT} and two instruments on INTEGRAL \citep{GCN_GRB150101B_INTEGRAL}.

Although \textit{Swift} BAT and \textit{Fermi} GBM have comparable sensitivities to SGRBs \citep{Burns_GBM_BAT}, GRB 150101B did not trigger the BAT instrument because the \textit{Swift} spacecraft was slewing at the time it occurred. 
% This slewing operation resulted in the detection of GRB 150101B through a ground analysis of slew data and subsequent non-standard data products compared to triggered GRBs. 
The initial circular \citep{GCN_GRB150101B_BAT} localized the burst to (RA, Dec) = (188.044, -10.956) and reported a single peak structure with a T90 of $\sim$18 ms. Two spectral fits were described: a blackbody with kT = (10 $\pm$ 2) keV and a power law with photon index (3.3 $\pm$ 1.5), exceptionally soft for a SGRB. Additionally, a significant spectral lag was reported, which is rare for SGRBs. With the knowledge of the analysis presented here, this lag may be due to the second component which was not independently identified in the BAT data. At the time of detection the lag contributed to an ambiguous characterization of the event and the initial BAT circular did not conclusively classify it as a GRB. The event does appear in the Third \textit{Swift} BAT GRB Catalog \citep{Swift_BAT_3rd_GRB_Lien_2016}, which reports a power law fit measured over the burst duration of 16 ms with an index of -1.5, which is more typical for SGRBs and consistent with the power-law index from the GBM catalog\footnote{\url{https://heasarc.gsfc.nasa.gov/W3Browse/fermi/fermigbrst.html}.}

\begin{figure*}
\centering
\includegraphics[width=1\textwidth]{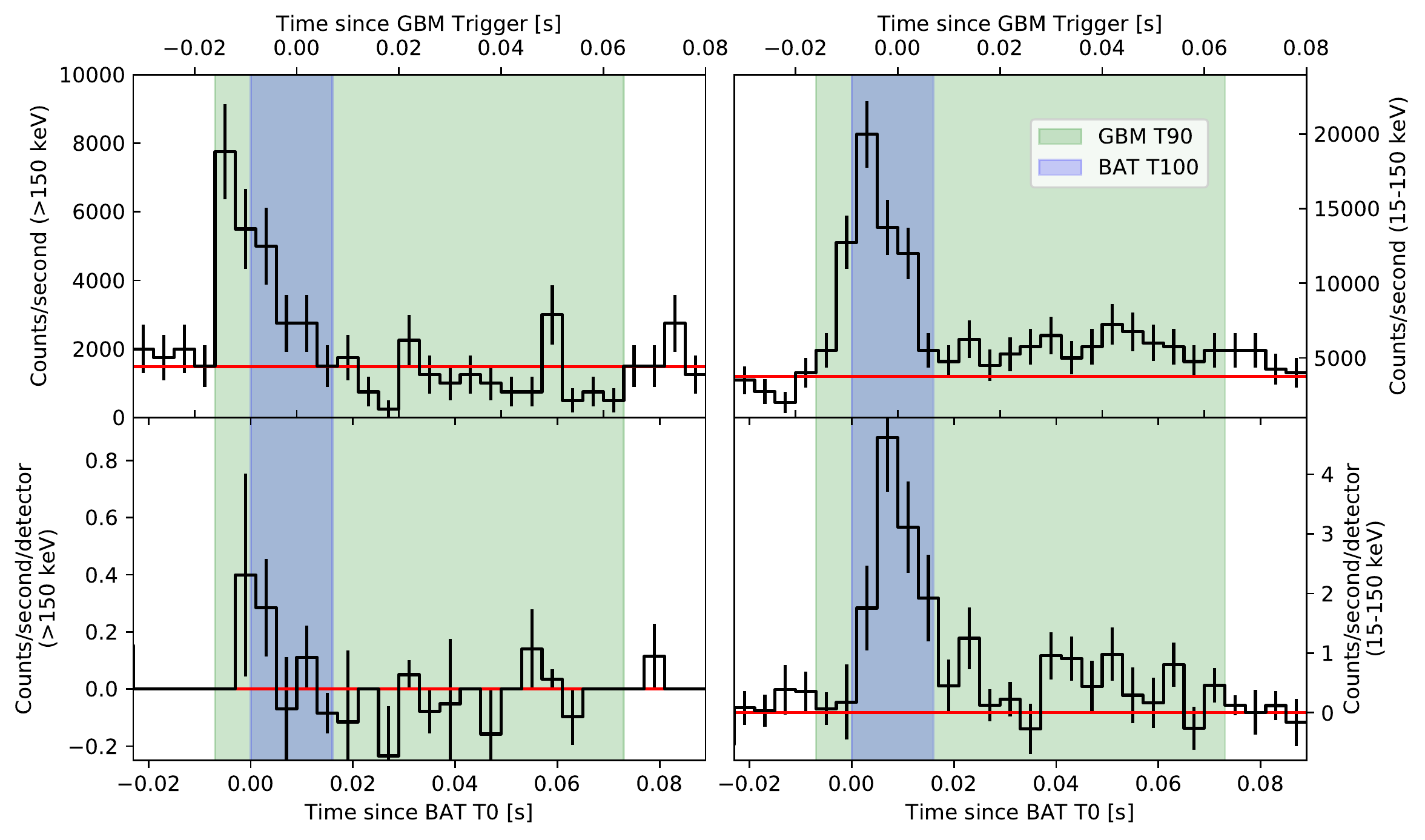}
	\caption{{The counts lightcurves for GRB 150101B as observed by \textit{Fermi} GBM (top) and \textit{Swift} BAT (bottom) in the 15-150 keV (right) and $>$150 keV energy ranges (left). Shown are 4 ms lightcurves in black and a background estimate in red. Time axis values are relative to the T$_0$ times of each instrument (GBM on top with T$_0$ defined as the trigger time, and BAT on bottom with T$_0$ defined as the start of the T100 interval since BAT did not trigger on-board). The lightcurves are aligned by correcting for light travel time between \textit{Fermi} and \textit{Swift} (a difference of 1.35 ms, with the burst arriving at \textit{Swift} first). The standard duration measure of each instrument are shown with the BAT T100 interval in blue and the GBM T$_{90}$ interval in green.}%{\it [Left]} 
\label{fig:LC_BAT_GBM}}
\end{figure*}

In \citet{Burns_GBM_BAT}, we previously investigated the observational differences of GRB 150101B as seen by BAT and GBM. We present here a fuller understanding, that is aided by Figure \ref{fig:LC_BAT_GBM}. The 16 ms interval from the BAT Catalog is not the same as the 16 ms of the main peak as observed by GBM. After accounting for the relative trigger times and the light travel time, the BAT 16 ms interval starts several ms after the GBM interval. There is evidence for both emission episodes in BAT. The BAT energy range is narrower at both the low and high ends compared to \textit{Fermi} GBM (15 -- 150 keV compared to 8 keV -- 40 MeV). As shown in Figure \ref{fig:GRB150101B_Flux} and Table \ref{tab:spectral_results}, the start of the main peak has a higher peak energy, higher energy flux, and lower photon flux than the end of the main peak, which may explain the lack of significant signal in BAT at the start of the burst. There is a hint of the hard emission above 150 keV in BAT, during the main peak identified by GBM. This emission is captured by the GBM measure of duration as it is performed in flux space. The BAT calculation of duration is performed in counts space, where the measure is dominated by the more numerous lower-energy counts. This is also a result of GBM having a larger effective area at higher energies than BAT. There is also evidence for the soft emission in BAT: the initial fits reported in \citet{GCN_GRB150101B_BAT} and the fits reported in \citet{GRB150101B_Troja} prefer a blackbody, consistent with the GBM observations of the soft tail. However, the soft tail was not independently identified in Swift BAT. The source position was at a good geometry for Fermi GBM, with 5 NaI and both BGO detectors having good views. The source position occurred at $\sim$40\% partial coding fraction for Swift BAT. The soft tail may be less significant in BAT due to the sensitivity of each instrument to the source position at trigger time and the higher low-energy limit of the BAT. The authors of \citet{GRB150101B_Troja} reach similar conclusions on the differences between the BAT and GBM observations of GRB 150101B (private communication).

In response to \citet{GRB150101B_Troja}, observations of GRB 150101B by the INTEGRAL spacecraft were recently reported \citep{GCN_GRB150101B_INTEGRAL}. The INTEGRAL team report marginal detections in SPI-ACS and IBIS-PICsIT. SPI-ACS data is 50 ms long, which is a factor of a few longer than the 16 ms main peak measured in GBM data. Despite this, a sharp spike is observed. As SPI-ACS has increased sensitivity at higher energies and no sensitivity below 75 keV this confirms the existence of spectrally hard emission in GRB 150101B.

GRB 150101B has high flux and a high peak energy and was within the field of view of the \textit{Fermi} Large Area Telescope (LAT) at event time, but there is no significant detection in the LAT. 

\end{document}